\documentclass[pra, 12 pt]{revtex4}
\usepackage[english]{babel}
\usepackage{amsmath}
\usepackage{epsfig}

\begin{document}

\title{Primordial plasma: influence of non-ideality on equilibrium radiation}

\author{S.A. Trigger$^{1,2}$, S.A. Maslov$^{1}$}
\address{Joint\, Institute\, for\, High\, Temperatures, Russian\, Academy\,
of\, Sciences, 13/19, Izhorskaia Str., Moscow\, 125412, Russia;\\
email:\, satron@mail.ru;\\
$^2$ Institut f\"ur Physik, Humboldt-Universit\"at zu Berlin,
Newtonstra{\ss}e 15, D-12489 Berlin, Germany}

\begin{abstract}

The asymptotic behavior of the equilibrium radiation in Maxwellian plasma is studied for the low--frequency region.  It is shown, that the equilibrium radiation can essentially deviate from the Planck law even for a weakly non-ideal plasma. The deviation from the Planck distribution is described by the transverse dielectric permittivity which takes into account both frequency and spatial dispersion. The influence of plasma non-ideality increases with the non-ideality parameter in the dielectric permittivity. The spectral energy distribution of the equilibrium radiation essentially changes in the region of observable frequencies. At asymptotically low frequencies, the equilibrium radiation  begins to exhibit power behavior instead the logarithmic one. The results indicate that the spectral energy distribution of radiation could be different then the Planck one in the primordial plasma of the early Universe. \\

PACS number(s) : 42.50.Ct, 52.25.Os, 52.40.Db
\end{abstract}

\maketitle

\section*{Introduction}

The spectral energy distribution of the equilibrium radiation established by M. Planck [1] corresponds to an idealized model of an absolutely black body, which exists in a cavity filled with radiation and bounded by an absolutely absorbing substance. It is assumed that the radiation is in thermodynamic equilibrium with the substance, although the effects of the interaction of photons with the substance bounding the cavity are not considered [2].

The practical implementation of the Planck distribution, as a rule, is associated with the consideration of a macroscopic body in thermal equilibrium with the "black" radiation surrounding it [2]. Great success has been achieved in solving this problem, which is directly related to Kirchhoff's law (see [3–-5] for more detail and reference therein). At the same time, lack of attention has been paid to the question of the spectral energy distribution of equilibrium radiation (SEDER) in a substance (see [6] and the literature cited there). The solution of this problem was mainly restricted to the analysis of transparency regions at low photon momentum. This approach seems to be limited, since it is clear from physical considerations that the effects of radiation absorption should be take into account to establish the thermodynamic equilibrium of radiation in matter. 

Recent investigations have been devoted to the sequential consideration of the effect of an absorbing plasma medium on the spectral energy density of equilibrium radiation (SEDER) in a substance [7–-9]. In these works, the consideration was carried out both for a completely equilibrium system of nonrelativistic charged particles and photons [7,8] in the quantum electrodynamic (QED) approximation, and based on a generalization of a more traditional approach using the fluctuation-dissipation theorem [9]. In [10, 11], based on [7, 8], the frequency-asymptotic behavior of SEDER was studied. It was found that at low frequencies the SEDER in a plasma medium has a logarithmic (integrable) singularity. In this case, the radiation energy remains finite. However, the works [10, 11] were based on the part of the SEDER that is related to the influence of plasma on the photon distribution function through the transverse permittivity of charged particles. At the same time, as was shown in [12, 13], there is one more contribution to the SEDER, which must be taken into account. This contribution is related to the interaction between intrinsic fluctuations of currents and fields and leads to a contribution of the same order as that considered earlier. Below, when calculating the low-frequency behavior of SEDER, we use the general approach developed in [12, 13]. The analytical asymptotic based on this approach changes the logarithmic low-frequency asymptotic singularity to the more singular, but still integrable one $1/\omega^{2/3}$ [14]. For simplicity, we restrict the consideration to radiation in an electron gas.

\section{General formulae for the SEDER in electron gas
}
According to [12,13], the full SEDER in plasma medium $e(\omega)$, which includes the zero fluctuations in plasma is the sum of two terms
\begin{eqnarray}
e(\omega)=e^{(1)}(\omega)+e^{(2)}(\omega)
 \label{F1}
\end{eqnarray}
\begin{eqnarray}
e^{(1)}(\omega)=\frac{V\hbar \omega^3}{\pi^2 c^3}\coth\left(\frac{\hbar\omega}{2T}\right)\frac{c^5}{\pi \omega}\int_0^\infty d k k^4 \frac{{\textrm {Im}}\varepsilon^{tr}(k,\omega)}{(\omega^2 {\textrm {Re}}\varepsilon^{tr}(k,\omega)-c^2 k^2)^2+\omega^4({\textrm {Im}}\varepsilon^{tr}(k,\omega))^2},
 \label{F2}
\end{eqnarray}

\begin{eqnarray}
e^{(2)}(\omega)=V
\sum_a \frac{\hbar\omega^2 \omega^2_{p,a}\coth \left(\frac{\hbar\omega}{2T}\right)}{\pi^3} \int_0^\infty d k k^2\frac{\textrm{Im} \varepsilon^{tr}(k,\omega)}{\mid\omega^2\varepsilon^{tr}(k,\omega)-k^2c^2\mid^2}\;\;
\label{F3}
\end{eqnarray}
Here $\varepsilon^{tr}(k,\omega)$ is the transverse dielectric permittivity (TDP) of non-relativistic plasma, which takes into account spatial and frequency dependence of electromagnetic field, as well as the arbitrary strong interaction between charged particle in the system. The value $\omega_{p,a}=\sqrt{ 4\pi n_a^2 z^2_a e^2/m_a}$ is the plasma frequency for the particles of species $a$.

As easy to see Eq.~(1), contains also the Planck distribution, which corresponds to the limit of negligible particle density.
We have stress that the zero fluctuations in plasma may in principle differ [15] from vacuum zero fluctuations, and the form of these fluctuations, strictly speaking, is not known. Both functions $e^{(1)}(\omega)$ and $e^{(2)}(\omega)$ are definitely positive for an arbitrary TDP form for arbitrary frequencies due to inclusion of zero fluctuations. For the low-frequency region considered in this paper, zero oscillations are negligible.

In this work, using the results of analytical and numerical calculations [14] for almost ideal plasma,  we consider the influence of non-ideality on the low--frequency behavior of the SEDER, when the Coulomb interaction parameter $\Gamma=e^2 n^{1/3}/ T$ increases. The particular form of the TDP $\varepsilon^{tr}(k,\omega)$ which we use for weakly non-ideal electron gas is written as
\begin{eqnarray}
{\textrm {Re}}\varepsilon^{tr}(X, W)=1-\frac{2 \Gamma_e \eta_e^{2/3}}{W^2}-\frac{ \Gamma_e \eta_e^{2/3}}{W^2}\left\{\left[\frac{W}{X^2}+ \frac{X^2}{2}\right]\times \right.\nonumber\\
\left.
\left[_1 F_1\left(1,\frac{3}{2};-\left(\frac{W}{2X}-\frac{X}{2}\right)^2\right)-_1 F_1\left(1,\frac{3}{2};-\left(\frac{W}{2X}+\frac{X}{2}\right)^2\right) \right]\right\} +
\frac{ \Gamma_e \eta_e^{2/3}}{W^2}\left \{\left[1+\frac{X^2}{2} \right]\times \right.\nonumber\\
\left.
\left[_1 F_1\left(1,\frac{3}{2};-\left(\frac{W}{2X}-\frac{X}{2}\right)^2\right)+_1 F_1\left(1,\frac{3}{2};-\left(\frac{W}{2X}+\frac{X}{2}\right)^2\right)\right]\right\} ,
\label{F4}
\end{eqnarray}
\begin{eqnarray}
{\textrm {Im}}\varepsilon^{tr}(X, W)=\frac{\sqrt\pi\, \Gamma_e \eta_e^{2/3}}{W^2} \left \{\frac{1}{X} +\frac{X}{2\pi}\right\} \left\{\exp\left[-\left(\frac{W}{2X}-\frac{X}{2}\right)^2\right]-\exp\left[-\left(\frac{W}{2X}+\frac{X}{2}\right)^2\right]\right\},
\label{F5}
\end{eqnarray}
where $\eta_e = n_e \Lambda_e^3$,  $W=\hbar\omega/T$, $X=k\Lambda_e/\sqrt{4\pi}$, $\Lambda_a=(2\pi\hbar^2/m_e T)^{1/2}$ is the thermal de Broglie wavelength for electrons and $_1F_1(\alpha, \beta, z)$ is the degenerate hypergeometric function.

\section{Asymptotical behavior of the SEDER and the Coulomb interaction influence
}

\begin{figure}[t]
\begin{center}
\includegraphics[width=3in]{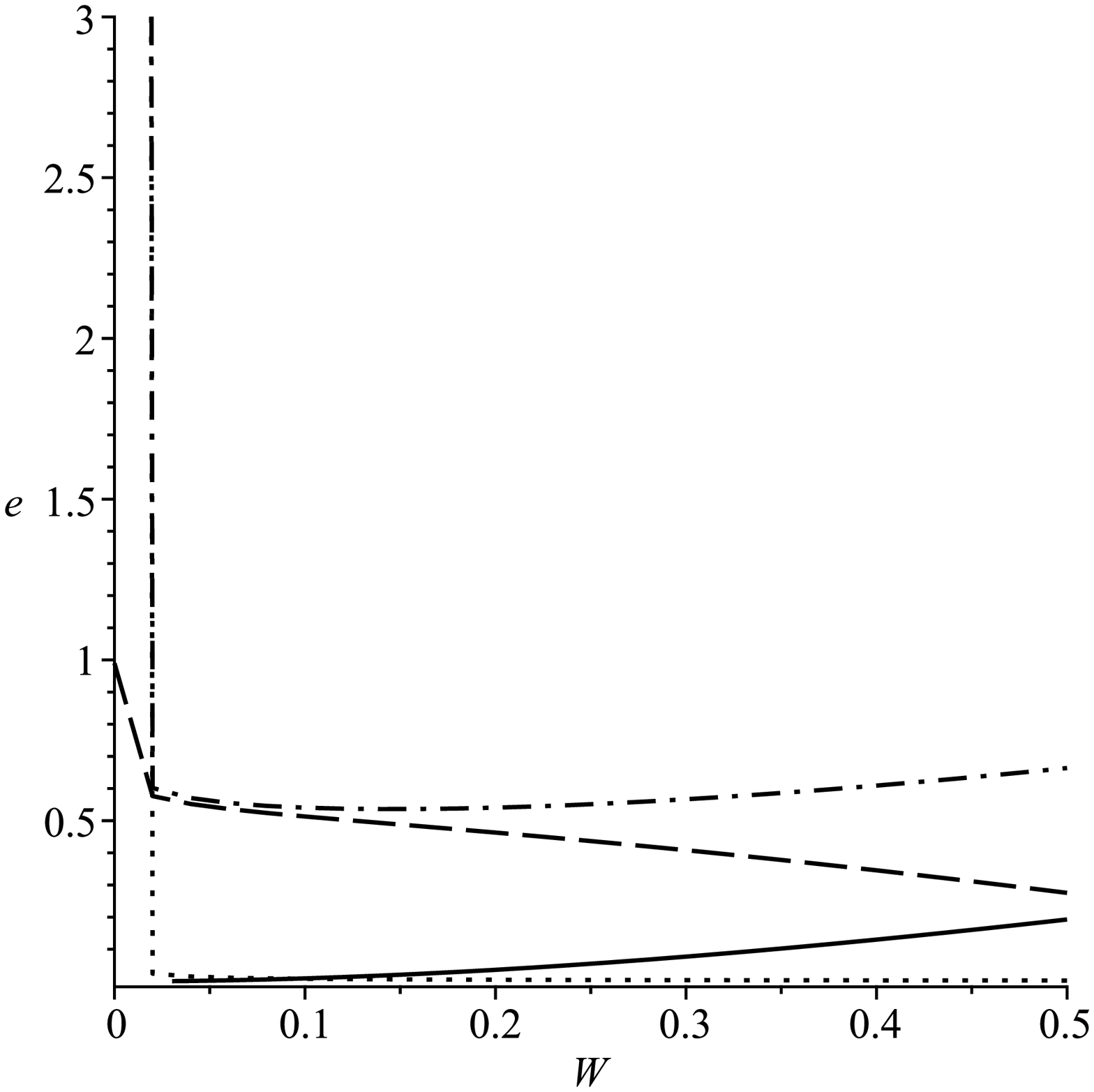}
\end{center}
\caption{\label{figure{1}} Different terms of the SEDER for $\eta =0.1$ and $\Gamma=0.02$. The Planck SEDER - (solid line); $e^{(1)}(W)$  (dashed line); $e^{(2)}(W)$  (dotted line); and the full SEDER in plasma  $e^{(1)}(W)+ e^{(2)}(W)$ (dash--dotted line). Calculations are valid for $W\ll 1$; however, formally, they are extended to the region $W\geq 0.5$.}
\end{figure}

As is known [10,11], the low-frequency asymptotic $W\ll 1$ of the part $e^{(1)}(\omega)$ of the SEDER has the logarithmic singularity
\begin{eqnarray}
e^{(1)}(\omega)\mid_{\omega\rightarrow 0}\rightarrow \frac{\sqrt 2\,\Gamma \eta^{2/3}}{3\sqrt{\pi\alpha}}\left[4\ln 4-3\gamma+\frac{6}{\pi}-2\ln(\alpha\sqrt\pi \Gamma\eta^{2/3}W)\right]\simeq 91,1 \,\eta^{1/3}\Gamma^2[28-2\ln(\frac{\eta^{4/3}W}{\Gamma})], \label{F6}
\end{eqnarray}
where $\gamma=0,577$ is the Euler's constant. The low-frequency asymptotic $W\ll 1$ of the part $e^{(2)}(\omega)$ of the SEDER [12,13] for electron gas has been recently analytically found [14] and reads
\begin{eqnarray}
e^{(2)}(\omega)\mid_{\omega\rightarrow 0}\rightarrow \frac{4\Gamma^2\eta^{4/3}\sqrt{2\pi\alpha}}{3\sqrt 3\, (\alpha\sqrt2\pi \Gamma\eta^{2/3}W)^{2/3}}\simeq 7.32
\frac{\Gamma^{5/3}\eta^{7/9}}{W^{2/3}},\label{F7}
\end{eqnarray}
where $\alpha\equiv Ry\, \eta^{2/3}/\pi m_e c^2 \Gamma$.

As easy to see in Figures 1 and 2, the Coulomb interaction increase leads to the increase in the SEDER for $W<0.5$. This value of the SEDER is essentially higher than the Planck one for small $W \ll 0.5$. In the region of frequencies $W\ll 0.1$, the singular behavior $\sim 1/W^{2/3}$ is dominant due to the term $e^{(2)}(W)$. For larger $W$, the crucial role is played by the logarithmic singularity of $e^{(1)}$. The Planck distribution is essential for the frequencies of the order of $W\simeq 0.3\div 0.5$ and even more, where the asymptotic relations (6) and (7) are strictly speaking invalid.

\begin{figure}[t]
\begin{center}
\includegraphics[width=3in]{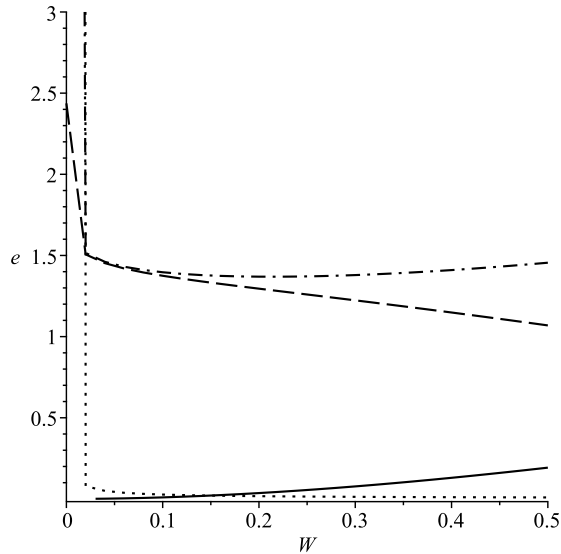}
\end{center}
\caption{\label{figure{2}} Different terms of the SEDER for $\eta =0.1$ and $\Gamma=0.03$. The Planck SEDER - (solid line); $e^{(1)}(W)$ (dashed line); $e^{(2)}(W)$ (dotted line); and the full SEDER in plasma $e^{(1)}(W)+ e^{(2)}(W)$ (dash-dotted line). Calculations are valid for $W\ll 1$; however, formally, they are extended to the region $W\geq 0.5$.}
\end{figure}

\section{Conclusions}

In conclusion, we emphasize that the above results take place for the homogeneous and isotropic Coulomb system that is in thermodynamic equilibrium with a thermostat at a given temperature $T$. This system contains charged particles and a quantized electromagnetic field which interact with each other. We consider the case of non-relativistic and non-degenerate charges under conditions when the account for quantum effects in TDP is fundamental for the convergence of the respective integrals at large wave vectors. In the above general expressions for the spectral energy density of the equilibrium radiation, these quantum effects are associated with the spatial and temporal dispersion of the transverse TDP. The explicit analytical expressions for the low-frequency asymptotical regimes for the SEDER are valid for plasma with a small interaction parameter $\Gamma < 1$. Since the results depend on the interaction parameter $\Gamma$ we can estimate the effect of the Coulomb interaction
on the SEDER. It is shown that an increase of the parameter $\Gamma$ leads to increase in the SEDER at low frequencies, where the role of plasma particles is dominant and the SEDER crucially differs in comparison with the Planck distribution.

In application to cosmic microwave background radiation (CMBR), the considered effects can be essential in observable radiation region. The maximal dimensionless frequency, according to the Planck distribution at $T \simeq 2,72$ K, is $W_{max}\simeq2,58$ and lies outside the asymptotical region of frequencies. However, for a fixed frequency, in the hot primordial plasma before the recombination epoch, the equilibrium radiation could be different from the Planck distribution, since the characteristic dimensionless frequencies $W_0$ are shifted to the low-frequency region. This difference can be essential for the description of the Universe evolution.

\section*{Acknowledgment}

The authors are thankful to A.M. Ignatov and A.G. Zagorodny for useful discussions.

\end{document}